\documentclass[sigconf]{acmart}

\AtBeginDocument{%
  \providecommand\BibTeX{{%
    \normalfont B\kern-0.5em{\scshape i\kern-0.25em b}\kern-0.8em\TeX}}}

\setcopyright{acmcopyright}
\copyrightyear{2019} 
\acmYear{2019} 
\acmConference[RecSys '19]{Thirteenth ACM Conference on Recommender Systems}{September 16--20, 2019}{Copenhagen, Denmark}
\acmBooktitle{Thirteenth ACM Conference on Recommender Systems (RecSys '19), September 16--20, 2019, Copenhagen, Denmark}\acmDOI{10.1145/3298689.3346966}
\acmISBN{978-1-4503-6243-6/19/09}

\begin{document}

\title{IRF: Interactive Recommendation through Dialogue}

\author{Oznur Alkan,  Massimiliano Mattetti, Elizabeth M. Daly, Adi Botea, Inge Vejsbjerg}
\email{{oalkan2, massimiliano.mattetti, elizabeth.daly, adibotea, ingevejs}@ie.ibm.com}
\affiliation{%
  \institution{IBM Research}
  \city{Dublin}
  \state{Ireland}
}

\renewcommand{\shortauthors}{Alkan, et al.}

\begin{abstract}
Recent research focuses beyond recommendation accuracy, towards human factors that influence the acceptance of recommendations, such as user satisfaction, trust, transparency and sense of control.
We present a generic interactive recommender framework that can 
add interaction functionalities to non-interactive recommender systems.
We take advantage of dialogue systems to interact with the user and we design a middleware layer to provide the interaction functions, such as
providing explanations for the recommendations, managing users' preferences learnt from dialogue, preference elicitation and 
refining recommendations based on learnt preferences.
\end{abstract}



\maketitle

\section{Introduction}
Most recommender systems offer very limited or no means to inform the recommender that its assumptions (e.g., about user preferences) 
are incorrect or outdated.
%to specify that preference information has become outdated.
%~\cite{jannach2016user}. 
However, users appreciate being more actively involved in the recommendation process,
and their feedback can lead to better recommendations.
%Enabling users to interact with the recommender could enable them to \textit{critique} specific features associated with the item, which in the end provides a much richer understanding of the users' preferences. 

%Wrapping an existing non-interactive algorithm into 
%an interactive layer 
%appears to be a valuable option, compared to
%throwing away existing algorithms and designing interactive recommenders from scratch, given that state-of-the-art recommender algorithms have reached a mature level (e.g., in terms of accurracy).
%
%On the other hand, many recommender algorithms employ advanced machine learning solutions that are not interpretable or actionable for the user. Additionally, a great deal of modelling and work may have been invested in the underlying algorithm, meaning service providers will be reluctant to forgo the advantages and accuracy of their existing solution for the sake of supporting additional user interaction mechanisms. 

We present a generic Interactive Recommender Framework (\textit{IRF}) that: \textit{1. can leverage an existing non-interactive recommender and turn it into an interactive recommender 2. implements the modes of interaction with the user.} To tackle the first problem, IRF is designed to be easily deployed on top of any existing non-interactive recommender, with a minimal configuration. Specifically, IRF can consume any recommender that aligns to the required API specifications, and the endpoints are designed based on \cite{Garcia2018}.
%\cite{RecSysWiki, Garcia2018}. 
To handle the second problem, IRF implements different types of interactions to acquire preferences from the user in a context where the user is motivated to give them, as well as to facilitate the exploration of the domain and the elicitation of the user's preferences. To achieve this, IRF implements the following interaction mechanisms: \textit{ presenting recommendations; explaining why an item is recommended; presenting the user profile; presenting item details; allowing users to provide their preferences on the feature or item level; and asking a preference elicitation question}. 
%With \textit{2, 3, 4, and 5}, users build up an understanding of the domain and the system's assumptions for their preferences as well as the relationship between the input to the system and output presented to the user. This will enable them to initiate a predictable and efficient interaction with the system so that they can meaningfully revise their input in order to improve recommendations \cite{Sinha2002}. Dialogue presents a unique opportunity for the solution to learn more about the user, which is the main motivation behind implementing interaction mechanism \textit{6}. This type of interaction allows the system to ask what it needs to learn, which leads the dialogue between the user and the recommender to resemble a more natural conversation where both sides can ask questions and respond. 
\section{System Overview}

%\subsection{Framework Requirements}
\textbf{Data.} Data stored in the backend have 4 main types: \textit{user\_profile, item\_profile, rec\_list}, and \textit{user\_preferences}. The first three are at the core of any recommender system and allow the configuration of required generic API endpoints. \textit{user\_preferences} is used within IRF to store the preferences for features learnt from the conversation. 
\textit{user\_profile} is required to contain a \textit{history} field, which is a list of \textit{<item, score, timestamp>} triples where \textit{score} and \textit{timestamp} are optional. It can contain further optional information depending on the underlying recommender service. If provided, they are sent to the recommender as part of the request. \textit{item\_profile} includes a list of features for the categories of the underlying domain. 
%For instance, possible movie categories can be genre, cast, director, etc., and genre features can be drama, action etc. \textit{rec\_list} is a list of \textit{<item, score, explanation>} triples, where explanations are optional. %Finally, the \textit{user\_preferences} stores a list of \textit{<category, feature, score>} triples for each user, score representing the calculated preference weight for that feature.% User preferences are stored in an internal database within IRF. 

\textbf{External Services.}
IRF relies on the following external services: \textit{1. Recommender Service 2. Item Data Service 3. User Data Service}. We minimize the assumptions about the external services. IRF expects the endpoints specified in Table \ref{table:externalservices}. These endpoints are designed following REST principles~
%\cite{RecSysWiki, Garcia2018}.
\cite{Garcia2018}.
% and patterns \cite{soawithrest}.
Some endpoints are optional, such as $user/update$, called if a notification should be sent to the user data service when the user provides a new preference.

\begin{table}
\tiny
\centering
\caption{External Services - API endpoint details}
\begin{tabular}{ p{2cm} p{0.5cm} p{1cm} p{1cm} p{2cm} }
 \hline 
 URL&Required&Parameters&Returns&Description\\
  \hline 
recommend/get/& required & user\_profile & rec\_list & returns recommendations\\\hline 
user/get/{uid}&required &user\_id & user\_profile & returns a user\_profile \\
user/update & optional&user\_profile& - & updates a user\_profile when a new item preference is mentioned\\\hline 
item/get/{iid}& required& item\_id & item\_profile & returns an item\_profile\\
item/desc/{iid}& optional& item\_id& item description text & returns a description text for the item\_profile \\\hline 
\end{tabular}
\label{table:externalservices}
\end{table}

%\subsection{Solution Components}
%\medskip
%\noindent
\textbf{Solution Components.} IRF has 2 main components: \textit{Dialogue Manager (DM)} and \textit{Middleware (MW)}. 

\textbf{\textit{DM}} is responsible for coordinating and managing the conversation with the user through 1. \textit{analyzing the user's utterances}, 2. \textit{calling the external services or MW based on the type of the interaction}, 3. \textit{presenting the final response} to the user. %, which is either a question to be asked or personalized recommendations, whenever appropriate. 
DM manages the conversation with a dialogue plan created with AI planning as described in \cite{Botea-et-al-DeepDial19}.  %, which enables IRF to easily support complex interactions. %Based on the dialogue plan, DM decides which action to take or which interaction to activate considering the current state of the dialogue that has been updated according to the latest user's utterance. 
It uses Watson Assistant (WA),\footnote{https://www.ibm.com/watson/developercloud/conversation.html} an existing service
%~\cite{W17-5522} 
that
assigns to every user utterance an \textit{intent} and zero or more \textit{entity--value} pairs. %For example, for an utterance such as \textit{"I do not like action movies"},  the intent is recognized as \texttt{\#negative-pref}.Also, word "action" is recognized as a value for an entity called \texttt{feature}.
The \textit{workspace} file, which includes all the \textit{entity--value} pairs should be provided to IRF to configure the WA. For response generation, a configurable \textit{messages file} is kept. The messages file saves sentence structures, which can possibly have place-holders for the inputs that will be set by the DM before presenting that response to the user. %For example, a fixed response string \textit{"These are the recommendations I found for you: \$recs"} is retrieved to form the response upon recommendation request. The value of \textit{\$recs} is filled by DM based on the response from the MW, and the finalized response string is presented to the user later on. 

%\textbf{Middleware (MW).}
\textbf{\textit{MW}} uses the following components to convert a non-interactive recommender into an interactive one. 

\textit{Post Processor (PP).}
The preferences learnt during conversation should be reflected back to the recommendations immediately. %
%This overcomes a limitation of many recommender systems which are unable to allow a user to impact the recommendations in real-time. For example, a standard collaborative filtering algorithm cannot consume learnt feature preferences. 
%To address this, post-processor reranks the recommendations received from the recommender using the learnt user preferences.
To achieve this, whenever a recommendation request is received from the user, the following actions are executed: 1. DM calls MW with $user\_profile$ and $user\_preferences$. 2. MW calls external recommender service with $user\_profile$ to receive an initial set of recommendations, 3. MW calls the post processor to rerank this initial list using the $user\_preferences$. Step 3 is performed by calculating a weighted similarity score between the $user\_preferences$ and each item within the recommendation list using cosine similarity measure. Final score is calculated by taking a weighted average of the initial recommender score and the preference similarity score. %Finally, the top \textit{n} recommendations, where \textit{n} is configurable within IRF, are returned back to DM. 

\textit{User Profiling and Preferences Manager (UPM).}
%Interactive recommendations need to be dynamic allowing the user to update preferences frequently and see the impact of these preferences. However, not all preferences mentioned by the user may reflect user's \textit{permanent} interests, as some may only be short-term or \textit{temporary} preferences reflecting some contextual requirements or the desire to explore what-if scenarios. 
%Therefore, the 
UPM differentiates between a \textit{temporary} and a \textit{permanent} user profile.
The \textit{temporary profile} amplifies the preferences learnt during the conversation and is created at the start of the session based on the \textit{permanent profile}. It gets updated whenever the user states new preferences for features or preference for an item during a conversation session. At the end of a session the temporary profile is merged with the permanent profile by replaying the individual preferences expressed in the dialogue so the newly learnt information still impacts the \textit{permanent profile} but to a lesser extent controlled by configured weights. The next time the \textit{permanent profile} is retrieved, all values are decayed based on how long it has been since the user profile was updated to reflect changes in the user's interests over time. 

\textit{Explanation Generation (EG).} EG generates a justification text for every recommended item for which an explanation is not already provided. It builds a \textit{TF-IDF model} \cite{contentbased2011} on top of the recommendation domain when the IRF is first initialized. %Once the \textit{TF-IDF model} is built, each item in the recommendation space is kept internally as a vector of feature weights based on their TF-IDF score. 
Furthermore, it builds an explanation through calculating the similarity between the 
$item\_profile$, $user\_profile$ and $user\_preferences$ vectors that are built based on the TF-IDF model. %It first builds a user profile vector based on the items that exist in the \textit{history} field of the $user\_profile$ and the corresponding item vectors. Afterwards, it calculates a similarity score for every feature that coexists in the $user\_profile$ vector, $user\_preferences$ and the $item\_profile$. During this similarity check, if a feature of an item vector exists both in $user\_profile$ and $user\_preferences$, the value within $user\_preferences$ is prioritized. 
As a result, explanation generation returns a list of common features between the item and the user profile and their associated similarity scores.%, which is later on used by DM to generate a proper explanation text.

%\paragraph{Presenting User Profiles}
Similarly, for  presenting the user profile, IRF uses the union of the preferences from \textit{user\_preferences} and the \textit{user\_profile} objects.%, where if the same feature exists in both, score within \textit{user\_preferences } is used. 
%The preferences that user likes and dislikes are formed and presented as separate lists to the user. Features with the top \textit{k} highest preference scores from both combined lists is presented, \textit{k} being configurable. If user profile data service provides a description text for \textit{user\_profile}, it is used directly to form the final response string together with the text formed from \textit{user\_preferences} object.

%\subsubsection{Question Generation}
%\label{ssec:preference-management}
\textit{Question Generation (QG).}
QG formulates a preference elicitation question to 
rank and filter the list of candidates.
We use a feature selection method based on \textit{Information Gain} \cite{Yang:1997:CSF:645526.657137}.
Assume that $k$ is the number of distinct features in the domain,
and each item is seen as a $k$-tuple.
We compute the information gain for each feature in the space of the recommended items, and return the one
with the highest value.

To set up IRF, the following files should
be updated: \textit{1. configuration file, 2. workspace file that contains the entity types and values and 3. messages file that includes the specific
utterances to use while responding
the end-user.} The configuration file configures the system properties
such as the details of the external services, the number of recommendations
to present to the user, etc. Workspace and the messages
files are used by DM, as described in previous paragraphs.

\section{System Demonstration}
% \begin{figure}
%   \begin{subfigure}[b]{0.45\columnwidth}
%    \frame{ \includegraphics[width=\textwidth]{figures/1.png}}
%     \caption{}
%     \label{fig:userinterface1}
%   \end{subfigure}
%   %
%    \begin{subfigure}[b]{0.45\columnwidth}
%    \frame{ \includegraphics[width=\textwidth]{figures/3.png}}
%     \caption{}
%     \label{fig:userinterface2}
%   \end{subfigure}
%   \label{fig:userinterface}
%   %
%   \caption{Screenshot of IRF Prototype Interface}
% \end{figure}

\begin{figure}
  \centering
%  \begin{tabular}{cc}
  \includegraphics[width=.23\textwidth]{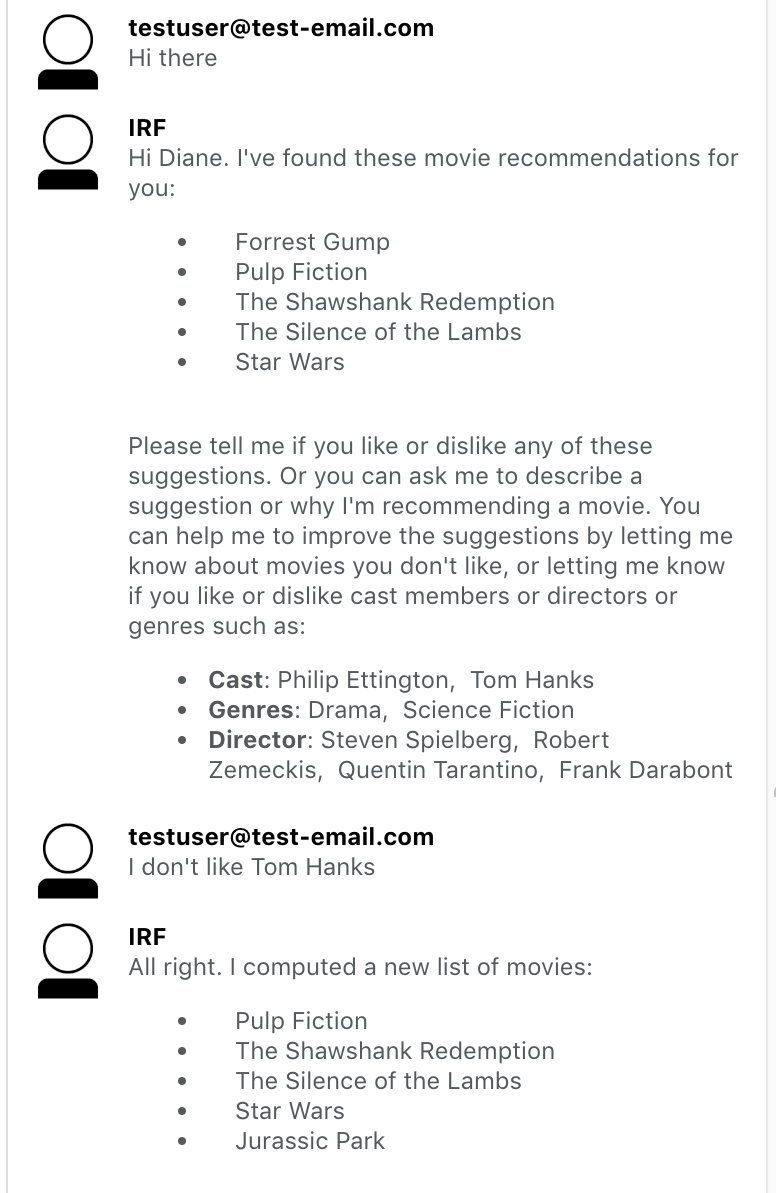}
  \includegraphics[width=.23\textwidth]{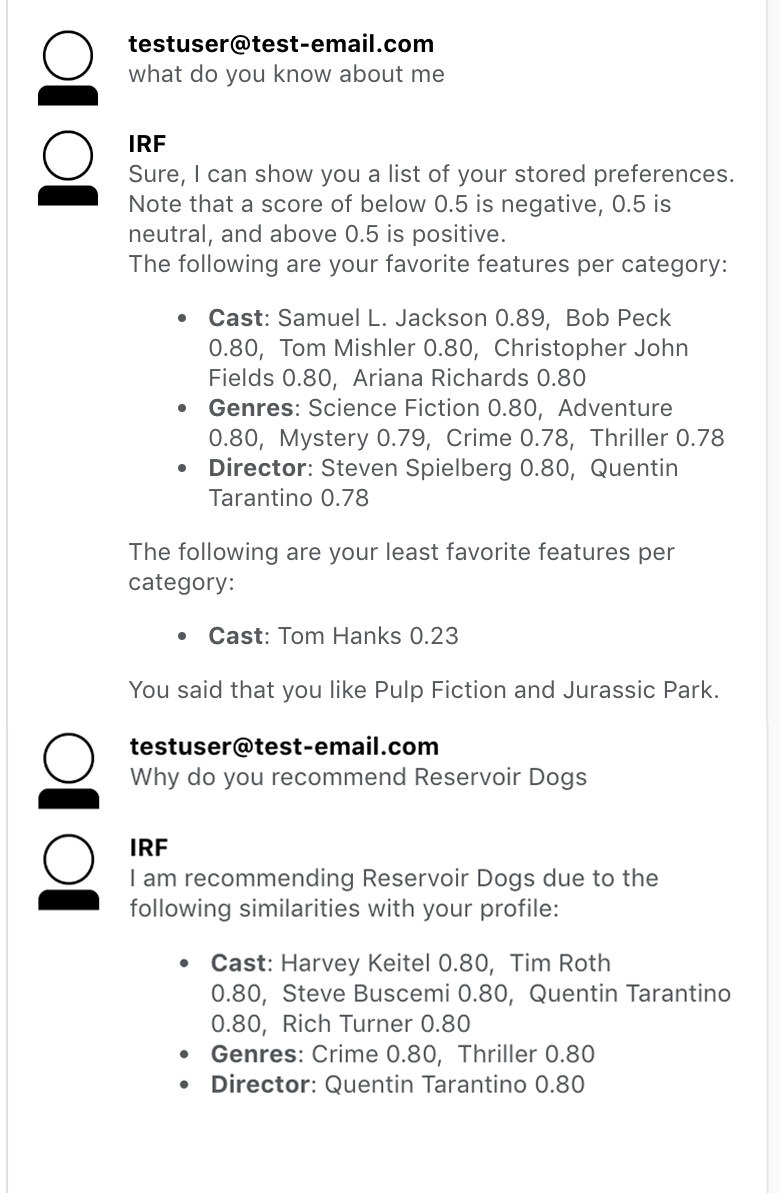}
%  \end{tabular}
  \caption{Sample Screenshots of the IRF prototype interface.}
  \label{fig:screenshots}
\end{figure}

To validate IRF's compatibility with different recommender algorithms and domains,we build two prototypes. One uses the MovieLens 100k\footnote{https://grouplens.org/datasets/movielens/latest/} dataset and a state of art content based recommender algorithm \cite{contentbased2011}. The second uses the Career Builder job recommendation\footnote{https://www.kaggle.com/c/job-recommendation} dataset with a sequential pattern mining based recommender algorithm \cite{Yap2012}.
From IRF's perspective, the only difference between the two systems are the setup files described earlier. Demonstration of the demo for interactive movie\footnote{https://www.youtube.com/watch?v=GwHyzLNVhNQ} and job\footnote{https://www.youtube.com/watch?v=wcgWAy5xoRw} recommenders are available on YouTube.  See Figure~\ref{fig:screenshots} for sample screenshots.
\section{Conclusion and Future Work}
We presented a framework for adding interaction capabilities to any existing non-interactive recommender solution which is independent of the underlying domain. %, and the recommender and the data services should comply with the API specifications. 
%In addition, the configuration file including some system configuration details together with the messages file and the workspace file should be provided. 
%We applied the framework to two different recommender algorithms for two different domains and the benefits are demonstrated with offline experiments and user studies. 
% THE FOLLOWING SOUNDS WEAK (AS IF THE CURRENT WORK IS NOT STRONG ENOUGH):
%The work has much room for further improvement.
A promising future direction is to design and test different preference elicitation strategies and allow more complex interactions with the user. %For example, dialogue can be used to clarify any conflicting information that exists in user profiles, which is especially the case for domains with many shared features between items. 

\bibliographystyle{ACM-Reference-Format}
\bibliography{references}


\begin{thebibliography}{5}

%%% the Web address.

\ifx \showCODEN    \undefined \def \showCODEN     #1{\unskip}     \fi
\ifx \showDOI      \undefined \def \showDOI       #1{#1}\fi
\ifx \showISBNx    \undefined \def \showISBNx     #1{\unskip}     \fi
\ifx \showISBNxiii \undefined \def \showISBNxiii  #1{\unskip}     \fi
\ifx \showISSN     \undefined \def \showISSN      #1{\unskip}     \fi
\ifx \showLCCN     \undefined \def \showLCCN      #1{\unskip}     \fi
\ifx \shownote     \undefined \def \shownote      #1{#1}          \fi
\ifx \showarticletitle \undefined \def \showarticletitle #1{#1}   \fi
\ifx \showURL      \undefined \def \showURL       {\relax}        \fi
% invisible to TeX
\providecommand\bibfield[2]{#2}
\providecommand\bibinfo[2]{#2}
\providecommand\natexlab[1]{#1}
\providecommand\showeprint[2][]{arXiv:#2}

\bibitem[\protect\citeauthoryear{Botea, Muise, Agarwal, Alkan, Bajgar, Daly,
  Kishimoto, Lastras, Marinescu, Ondrej, Pedemonte, and Vodolan}{Botea
  et~al\mbox{.}}{2019}]%
        {Botea-et-al-DeepDial19}
\bibfield{author}{\bibinfo{person}{Adi Botea}, \bibinfo{person}{Christian
  Muise}, \bibinfo{person}{Shubham Agarwal}, \bibinfo{person}{Oznur Alkan},
  \bibinfo{person}{Ondrej Bajgar}, \bibinfo{person}{Elizabeth Daly},
  \bibinfo{person}{Akihiro Kishimoto}, \bibinfo{person}{Luis Lastras},
  \bibinfo{person}{Radu Marinescu}, \bibinfo{person}{Josef Ondrej},
  \bibinfo{person}{Pablo Pedemonte}, {and} \bibinfo{person}{Miroslav Vodolan}.}
  \bibinfo{year}{2019}\natexlab{}.
\newblock \showarticletitle{Generating Dialogue Agents via Automated Planning}.
  In \bibinfo{booktitle}{\emph{The Second AAAI Workshop on Reasoning and
  Learning for Human-Machine Dialogues (DEEP-DIAL 2019)}}.
\newblock


\bibitem[\protect\citeauthoryear{Garc\'{\i}a and Bellog\'{\i}n}{Garc\'{\i}a and
  Bellog\'{\i}n}{2018}]%
        {Garcia2018}
\bibfield{author}{\bibinfo{person}{Iv\'{a}n Garc\'{\i}a} {and}
  \bibinfo{person}{Alejandro Bellog\'{\i}n}.} \bibinfo{year}{2018}\natexlab{}.
\newblock \showarticletitle{Towards an Open, Collaborative REST API for
  Recommender Systems}. In \bibinfo{booktitle}{\emph{Proceedings of the 12th
  ACM Conference on Recommender Systems}} \emph{(\bibinfo{series}{RecSys
  '18})}. \bibinfo{publisher}{ACM}, \bibinfo{pages}{504--505}.
\newblock
\showISBNx{978-1-4503-5901-6}


\bibitem[\protect\citeauthoryear{Lops, de~Gemmis, and Semeraro}{Lops
  et~al\mbox{.}}{2011}]%
        {contentbased2011}
\bibfield{author}{\bibinfo{person}{Pasquale Lops}, \bibinfo{person}{Marco de
  Gemmis}, {and} \bibinfo{person}{Giovanni Semeraro}.}
  \bibinfo{year}{2011}\natexlab{}.
\newblock \bibinfo{booktitle}{\emph{Content-based Recommender Systems: State of
  the Art and Trends}}.
\newblock \bibinfo{pages}{73--105}.
\newblock


\bibitem[\protect\citeauthoryear{Yang and Pedersen}{Yang and Pedersen}{1997}]%
        {Yang:1997:CSF:645526.657137}
\bibfield{author}{\bibinfo{person}{Yiming Yang} {and} \bibinfo{person}{Jan~O.
  Pedersen}.} \bibinfo{year}{1997}\natexlab{}.
\newblock \showarticletitle{A Comparative Study on Feature Selection in Text
  Categorization}. In \bibinfo{booktitle}{\emph{Proceedings of the Fourteenth
  International Conference on Machine Learning}} \emph{(\bibinfo{series}{ICML
  ’97})}. \bibinfo{pages}{412--420}.
\newblock


\bibitem[\protect\citeauthoryear{Yap, Li, and Yu}{Yap et~al\mbox{.}}{2012}]%
        {Yap2012}
\bibfield{author}{\bibinfo{person}{Ghim-Eng Yap}, \bibinfo{person}{Xiao-Li Li},
  {and} \bibinfo{person}{Philip~S. Yu}.} \bibinfo{year}{2012}\natexlab{}.
\newblock \showarticletitle{Effective Next-items Recommendation via
  Personalized Sequential Pattern Mining}. In
  \bibinfo{booktitle}{\emph{Proceedings of the International Conference on
  Database Systems for Advanced Applications}}. \bibinfo{pages}{48--64}.
\newblock


\end{thebibliography}

\appendix

\end{document}